# Development of ultrafast four-dimensional precession electron diffraction

Toshiya Shiratori[A,*], Jumpei Koga[A,*],

Takahiro Shimojima[B], Kyoko Ishizaka[A,B], and Asuka Nakamura[B,†]

*These authors equally contributed to this work.

† corresponding author. Email: asuka.nakamura@riken.jp

[A]Quantum-Phase Electronics Center and Department of Applied Physics, The University of Tokyo, Hongo, Tokyo 113-8656, Japan

[B] RIKEN Center for Emergent Matter Science (CEMS), Wako, Saitama, 351-0198, Japan

**Abstract**

Ultrafast electron diffraction/microscopy technique enables us to investigate the nonequilibrium dynamics of crystal structures in the femtosecond-nanosecond time domain. However, the electron diffraction intensities are in general extremely sensitive to the excitation errors (i.e., deviation from the Bragg condition) and the dynamical effects, which had prevented us from quantitatively discussing the crystal structure dynamics. Here, we develop a four-dimensional precession electron diffraction (4D-PED) system by which time ($t$) and electron-incident-angle ($\phi$) dependences of electron diffraction patterns ($q_x, q_y$) are recorded. Nonequilibrium crystal structure refinement on VTe$_2$ demonstrates that the ultrafast change in the crystal structure can be quantitatively determined from 4D-PED. We further perform the analysis of the $\phi$ dependence, from which we can qualitatively estimate the change in the reciprocal lattice vector parallel to the optical axis. These results show the capability of the 4D-PED method for the quantitative investigation of ultrafast crystal structural dynamics.

# 1. Introduction

The development of the time-resolved ultrafast electron diffraction/microscopy technique has opened a new pathway to investigate the photoinduced dynamics of materials. In the pioneering work by A. H. Zewail *et al.* [1,2], ultrafast electron diffraction and microscopy have been developed by combining the optical pump-probe technique with a photoelectron pulse. This method enables us to perform time-dependent electron diffraction/microscopy measurements in the femtosecond-nanosecond time domain. For example, Lorentz transmission electron microscopy [3–6], photoinduced near-field electron microscopy [7–9], and convergent-beam electron diffraction measurements [10,11] have been used to detect the dynamics of mesoscopic physical quantities such as magnetization, permittivity, and strain. Furthermore, combining four-dimensional scanning transmission electron microscopy (STEM) and pulsed electron (i.e., 5D-STEM) realize the quantitative spatiotemporal mapping of mesoscopic physical quantities such as time-dependent strain and magnetic field [12,13].

In particular, ultrafast electron diffraction is essential in investigating the crystal structure dynamics. It has revealed various phenomena, such as photoacoustic generation [14–20] and photoinduced structural phase transitions [19,21–28]. Compared with ultrafast time-resolved x-ray diffraction that requires free-electron laser facilities, ultrafast electron diffraction has an advantage in the beam time, which enables the variety of systematic measurements that require long data acquisition time. However, quantitative evaluation of crystal structure dynamics (i.e., change in atomic coordinates) from electron diffraction intensities has been difficult because of its sensitivity to the excitation error and dynamical effects of the Bragg diffraction [29,30]. As a result, discussion on the crystal structure dynamics based on the ultrafast electron diffraction intensity has been limited to determining the timescale and qualitative motion of atoms. In the normal (static) electron diffraction, this problem had been solved by the precession electron diffraction (PED) method [30,31]. In PED, the incident angle of the electron beam is continuously changed around the optical axis during the data acquisition, equivalent to the precession of the sample to a fixed electron beam parallel to the optical axis. Such a measurement significantly suppresses the effect of the excitation error and dynamic scattering, and can be used for quantitative structure refinement. Based on PED measurements, structural refinement on various solid materials had been achieved in the past decades [29,32–39], suggesting the validity of this method.

In this study, we develop an ultrafast PED that combines the pump-probe method with PED using the ultrafast transmission electron microscopy setup. Utilizing a pixelated detector and an ultrafast PED system enables us to obtain the diffraction patterns ($q_x$, $q_y$) for each electron-incident-angle $\phi$ and time $t$, which we refer to as four-dimensional precession electron diffraction (4D-PED). First, we verify the quantitative agreement between the PED intensity and the dynamical diffraction simulation obtained from high-quality single crystalline silicon. Second, we observe the change in PED intensity caused by photoinduced crystal structure dynamics in $VTe_2$ and demonstrate nonequilibrium crystal structure refinement in the picosecond scale. Third, electron-incident-angle $\phi$ dependent profile reveals the time-dependent change in the Bragg condition. With this we can determine the change in the reciprocal lattice vector parallel to the optical axis, which is difficult to probe in the conventional electron diffraction method. These results demonstrate the capability of the 4D-PED method for quantitative investigation of ultrafast crystal structure dynamics.

## 2. Development of ultrafast precession electron diffraction

### 2.1. Setup of measurement system

Fig. 1(a) shows a schematic of the 4D-PED measurement system developed in this study. A combination of femtosecond laser and transmission electron microscope enables us to perform ultrafast electron microscopy as described in the literature [19]. The sample is excited by a pump optical pulse and probed by an electron pulse, as shown in

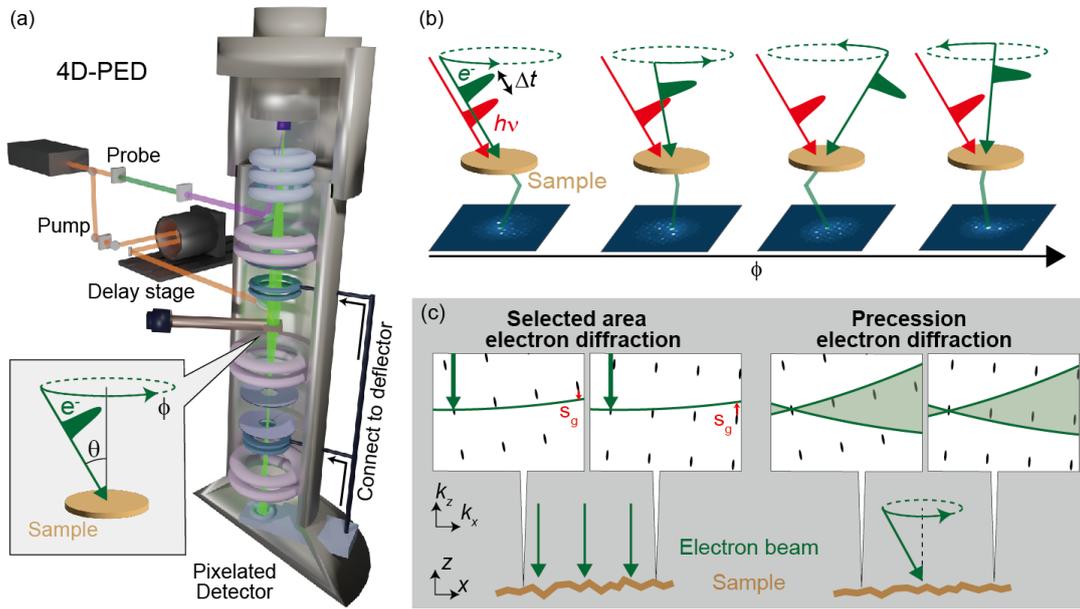

**Fig. 1. Experimental setup.** (a) Experimental setup for the 4D-PED system. (b) Schematic of 4D-PED data acquisition. The red pulse with $h\nu$ denotes the optical pulse, whereas the green pulse indicates the electron pulse. (c) Schematics of sample-position-dependent diffraction condition for selected-area electron diffraction (SAED) and PED. The upper panels show the reciprocal space. The green arrow and curve in the reciprocal space denote the incident vector of the probe electron and Ewald sphere, respectively. The black ellipse dots are the reciprocal lattice points that are rotated at different sample positions by unintended and local bending of the sample. PED suppresses the effects of non-systematic excitation error ($\mathbf{s_g}$) distribution by sweeping reciprocal space with respect to $\mathbf{s_g}$ compared to SAED.

Fig. 1(b). In the 4D-PED, the electron beam is tilted off the optical axis by a precession angle $\theta$ and precessed with an angle $\phi$ around the optical axis [Fig. 1(a,b)]. The diffracted electrons are tilted back so that the direct beam position in the diffraction pattern does not change with varying $\phi$ (de-scan). Conventionally, the precession of the electron beam is controlled by a dedicated external device, which provides continuous beam tilt and de-scan. We controlled the incident-angle $(\theta, \phi)$ of the electron beam and pixelated detector via software in the present case. This enables us to record the diffraction patterns for each $\phi$, similar to Ref. [40]. As a result of $\phi$- and time-resolved PED measurement (i.e., 4D-PED), we can obtain the four-dimensional diffraction intensity $I(q_x, q_y, \phi, t)$, where $q_x, q_y$ are the scattering vector components perpendicular to the optical axis.

We note that integrating $I(q_x, q_y, \phi, t)$ along $\phi$ axis significantly reduces the effect of excitation error and dynamical effect. For example, if the sample is unintentionally bent in the region of interest [Fig. 1(c)], it causes the non-systematic spatial distribution of excitation error $\mathbf{s_g}$ ($\mathbf{g}$ is a reciprocal lattice vector). Combined with the strong dynamical effect of electron diffraction, the quantitative discussion of the crystal structure based on the selected-area diffraction (SAED) intensities is difficult. Integrating the excitation error and off-zone axes incidence of the electron beam in PED significantly reduces the uneven diffracted intensities caused by the excitation error and dynamical effect. It thus enables us to measure the diffraction intensity that can be used for quantitative structural analysis.

**2.2. Dataset obtained from 4D-PED**

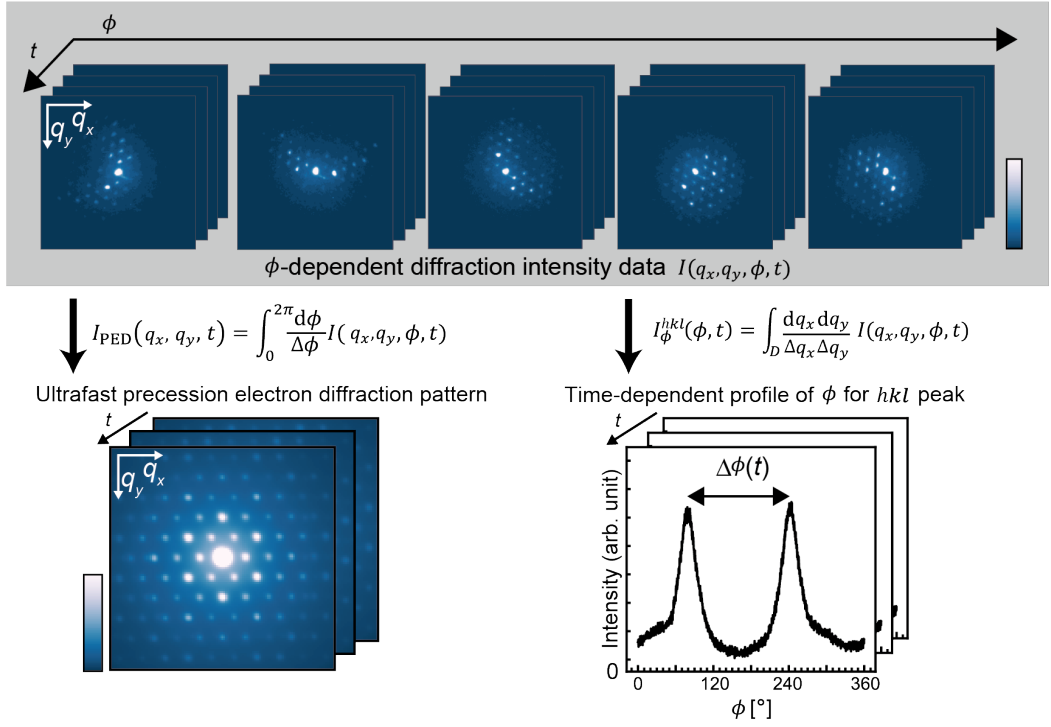

**Fig. 2. Dataset obtained from 4D-PED.** $\phi$- and time-resolved 4-dimensional data sets $I(q_x, q_y, \phi, t)$ by 4D-PED (top). Time-resolved PED intensities $I_{\text{PED}}(q_x, q_y, t)$ are obtained by integrating $I(q_x, q_y, \phi, t)$ with respect to $\phi$ (bottom left). Time-resolved $\phi$ profile of $hkl$ diffraction intensity $I_\phi^{hkl}(\phi, t)$ are taken by integrating $I(q_x, q_y, \phi, t)$ around the diffraction peak (bottom right).

From the 4D-PED, we obtain the time- and $\phi$-dependent diffraction intensity data $I(q_x, q_y, \phi, t)$ as displayed in the upper panel of Fig. 2. By $\phi$-integration of this data from 0 to $2\pi$, we obtain the time-dependent PED intensity $I_{\text{PED}}(q_x, q_y, t)$ as shown in the bottom left panel in Fig. 2. In general, PED intensity depends on the sample thickness, lattice parameters, atomic coordinates, and atomic displacement parameters, as well as the experimental conditions ($\theta$ and acceleration voltage). Therefore, it is possible to determine these crystal structural parameters and the sample thickness by fitting $I_{\text{PED}}(q_x, q_y, t)$ with the kinematical/dynamical diffraction intensity simulation as demonstrated by previous PED studies [29,32–39].

In addition, the $q_x, q_y$ integration of $I(q_x, q_y, \phi, t)$ around a specific diffraction spot $hkl$ gives $I_\phi^{hkl}(\phi, t)$ as depicted in the bottom right panel in Fig. 2, which provides information on the reciprocal lattice vector parallel to the optical axis. As shown in Fig. 2 bottom right panel, $I_\phi^{hkl}(\phi, t)$ has two peaks, where the $hkl$ diffraction satisfies the Bragg condition. In the 4D-PED setup, the center of the Ewald sphere is expressed as $\mathbf{K}_0 = K_0(\sin\theta\cos\phi, \sin\theta\sin\phi, \cos\theta)$ in the reciprocal space $(k_x, k_y, k_z)$. Here, we define $K_0 = |\mathbf{K}_0|$ and $k_z$ as parallel to the optical axis. Using reciprocal lattice vector $\mathbf{g}^{hkl}$ and solving Bragg condition $|\mathbf{K}_0 - \mathbf{g}^{hkl}| = K_0$, two solutions for $\phi(=\phi_1, \phi_2)$ are obtained. Then, we can calculate the difference between these two solutions, $\Delta\phi_{hkl} = \phi_1 - \phi_2$ as

$$\cos\frac{\Delta\phi_{hkl}}{2} = \frac{|\mathbf{g}^{hkl}|}{2K_0 \sin\theta \cos\delta_{hkl}} - \frac{\tan\delta_{hkl}}{\tan\theta}, \qquad (1)$$

where $\tan\delta_{hkl} = g_z^{hkl}/\sqrt{(g_x^{hkl})^2 + (g_y^{hkl})^2}$ and $\mathbf{g}^{hkl} = (g_x^{hkl}, g_y^{hkl}, g_z^{hkl})$ in reciprocal space. Therefore, experimentally obtained $\Delta\phi_{hkl}(t)$ can be used to estimate $\mathbf{g}^{hkl}$. It is noted that Eq. 1 can be primarily utilized to obtain $g_z^{hkl}$ because $g_x^{hkl}$ and $g_y^{hkl}$ can be easily evaluated from peak positions in the diffraction pattern. Quantitative estimation of $g_z^{hkl}$ enables us to obtain information on lattice parameters that cannot be given by conventional PED and SAED because these conventional methods provide no information for the components parallel to the optical axis $g_z^{hkl}$.

### 2.3. Experimental setup

The detailed setup of an ultrafast electron microscopy system is described elsewhere [19]. As a pixelated detector, we use MerlinEM (Quantum Detectors). In this study, we use the 12.5 kHz pump light pulse with a fluence of 2.1 $mJ/cm^2$. The diameter of the pump laser spot was set around 100 μm, much larger than the sample size (< 10 μm). We used the condenser aperture of 150 μm. The field of view was about 1 μm, which was enough smaller than the pump spot size. We set the precession angle $\theta$ as 3° (2°) for silicon (VTe$_2$). The number of divisions for $\phi$ direction was 1800. As a result, we obtained $I(q_x, q_y, \phi)$ for respective time as $256 \times 256 \times 1800$ three-dimensional data.

### 2.4. Data analysis and simulations

For all data, we shift the diffraction patterns at respective $\phi$ and $t$ to match the direct beam position at the center of the image because the direct beam position slightly varies with $\phi$ due to the incomplete de-scan. To extract the experimental diffraction intensities of $hkl$ diffraction $I_{\exp}^{hkl}(t)$ for the crystal structure refinement, we translate $I_{\text{PED}}(q_x, q_y, t)$ to those in the polar coordinate $I_{\text{PED}}(q_r, q_\theta, t)$ by linear interpolation, where the origin of $q_r, q_\theta$ is taken at the direct spot position. We subtracted the background intensity from $I_{\text{PED}}(q_r, q_\theta, t)$ before evaluating the intensities of the respective diffraction peaks $I_{\exp}^{hkl}$. For this purpose, we assume that the background intensity depends only on $q_r$ (i.e., isotropic in the reciprocal space), and the background intensity at respective $q_r$ and $t$ was estimated as 10 percentiles of $I_{\text{PED}}(q_r, q_\theta, t)$ along $q_\theta$ direction. After the background subtraction, we integrated $I_{\text{PED}}(q_r, q_\theta, t)$ along $q_r$ direction around the respective $hkl$ diffraction and estimated the diffraction intensity $I_{\exp}^{hkl}(t)$ by fitting it with a Gaussian along $q_\theta$ direction. All data analyses were performed on the Python-based multi-dimensional data analysis platform *lys* [41].

The dynamical (multislice) electron diffraction simulations of silicon and VTe$_2$ were performed following Ref. [42]. For the simulations, the wave function was sampled with $128 \times 128$ ($256 \times 64$) points along the $a$ and $b$ axes of silicon (VTe$_2$). The slice thickness was set identical to the unit cell size perpendicular to $a$ and $b$ for silicon and VTe$_2$. The electron beam incidence was set perpendicular to the $a$ and $b$ axes (It is noted that this is not exactly parallel to the $c$ axis in monoclinic VTe$_2$). For the SAED simulation, we performed a single multislice simulation by setting the electron beam parallel to the optical axis. For the PED intensity calculation, the 1800 multislice simulations were performed by changing $\phi$ from 0 to $2\pi$ and were averaged to simulate the PED results.

## 3. Results
### 3.1. Verification of precession electron diffraction system

To begin with, we verify our experimental system by comparing the selected area electron diffraction (SAED) pattern with that of PED. Figure 3(a) shows the SAED pattern of silicon (001). The four-fold symmetric diffraction pattern reflects the cubic crystal structure of single crystalline silicon. Figure 3(b) shows the intensity profile along the red line in Fig. 3(a) (red curve) and the SAED simulation based on dynamical diffraction theory (blue dashed curve). Considering the cubic symmetry of the silicon, experimental data should be ideally symmetric with respect to $q_x = 0$. Nonetheless, the observed diffraction intensity profile in Fig. 3(b) is asymmetric. In addition, the experiment and simulation show the significant difference in the $k_x$-dependent diffraction intensities. As described in Sec. 2.1, these mismatches should be arising from the excitation error and dynamical effect of Bragg diffractions, which prevents us from experimentally extracting the information on the crystal structure. On the other hand, the PED pattern in Fig. 3(c) significantly differs from the SAED pattern. This can be confirmed by the intensity profile [Fig. 3(d)] that is almost symmetric with respect to $q_x = 0$. At the same time, the experimental data are similar to the simulation. This result demonstrates that applying 4D-PED dramatically improves the diffraction pattern, which can be directly compared with the simulations.

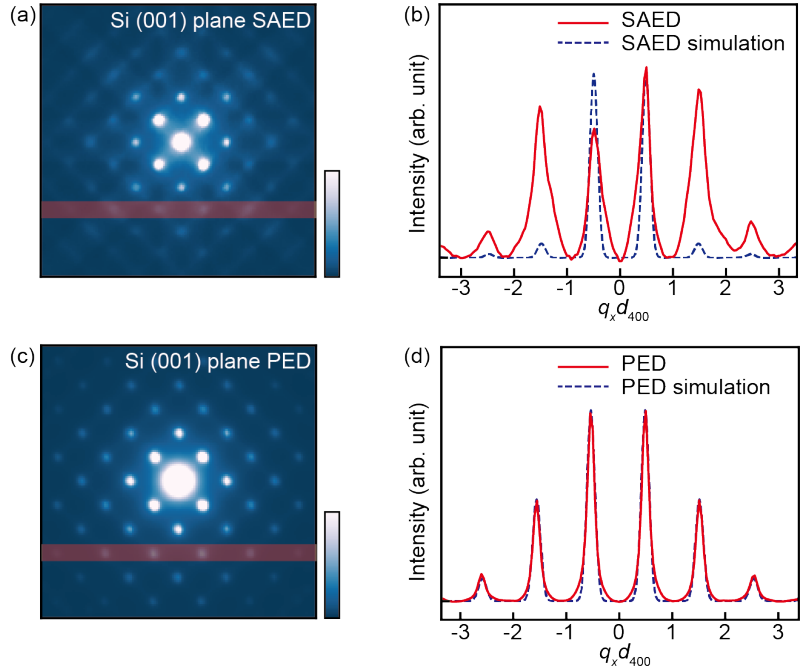

**Fig. 3. Verification of PED.** (a) SAED diffraction pattern of silicon (001) plane. (b) Comparison between experimental (solid red curve) and simulated (blue dashed curve) SAED intensity along with the horizontal red line in a. (c) PED diffraction pattern of silicon (001) plane. (b) Comparison between experimental (solid red curve) and simulated (dashed blue curve) PED intensity along the horizontal red line in c.

## 3.2. Demonstration of ultrafast precession electron diffraction

Hereafter, we focus on the experimental results of VTe$_2$. This is a suitable system for demonstrating 4D-PED since its photoinduced crystal structural dynamics has been well studied recently [18,19,43–45]. Figures 4(a-c) show the PED pattern and crystal structure of VTe$_2$ at room temperature [46,47]. In this paper, we use the definition of $a, b,$ and $c$ axes shown in Figs. 4(b,c) to describe the Miller indices. As the previous research [19] revealed, VTe$_2$ shows the photoinduced dissolution of the V-V ribbon chain structure [Fig. 4(b)] within 2 ps after the optical pump excitation at room temperature. Moreover, the angle between the $a$ and $c$ axes [$\beta$ shown in Fig. 4(c)] also changes from 90.37° towards 90° within several tens of picoseconds. The two-fold symmetric PED pattern in Fig. 4(a) indicates the zone axis incidence of the electron beam. Under such a condition, ideally, the time dependence of the diffraction intensity should be identical for $hkl$ and $h\bar{k}l$, since the crystal structure of VTe$_2$ (space group $C2/m$) has a mirror plane perpendicular to the $b$ axis.

We compare the experimental results of conventional ultrafast SAED with 4D-PED to demonstrate the advantage of 4D-PED. We observed a change in ultrafast PED intensity $I_{\text{PED}}(q_x, q_y, t)$ at several Bragg spots as indicated by $\Delta I = I(t) - I(t < 0)$ at $t = 50$ ps in Fig. 4(d). Such time-dependent Bragg diffraction intensities have been interpreted as the change in crystal structure in many previous ultrafast electron diffraction researches [19,21,22]. Although $\Delta I$ of ultrafast PED is symmetric with respect to the mirror plane indicated by the black dashed line as expected, that of ultrafast SAED in Fig. 4(e) does not satisfy this condition. Indeed, the time profiles of 6 2 0 and 6 $\bar{2}$ 0 diffraction intensities for ultrafast PED in Fig. 4(f) mostly overlap, while SAED intensities in Fig. 4(g) strongly deviate from each other. These results indicate that ultrafast PED intensities satisfy the symmetry requirement and offer clues for quantitative interpretations. For example, 6 2 0 and 6 $\bar{2}$ 0 ultrafast PED intensities suddenly drop at $t \sim 0$, indicating the dissolution of the V-V ribbon chain structure observed in Ref. [19]. Such a change in the crystal structure can also describe the symmetric increase in 18 0 0 and $\overline{18}$ 0 0 diffractions just after the photoexcitation ($t <$ 6 ps) [Fig. 4(h)]. For 18 0 0 and $\overline{18}$ 0 0 diffraction, asymmetric change in diffraction intensities is also observed in a longer timescale. In the present case, this can be explained by the change in $\beta$, which should significantly affect $\mathbf{s_g}$ asymmetrically for $hkl$ and $\bar{h}kl$ as discussed in Ref. [19]. Indeed, a change in $\beta$ is observed in this timescale, as later discussed in Sec. 3.4. Conversely, the conventional ultrafast SAED results in Figs. 4(g,i) do not satisfy the symmetric constraints (i.e. mirror symmetry), which prevents us from even discussing the qualitative change in the crystal structure from diffraction intensity.

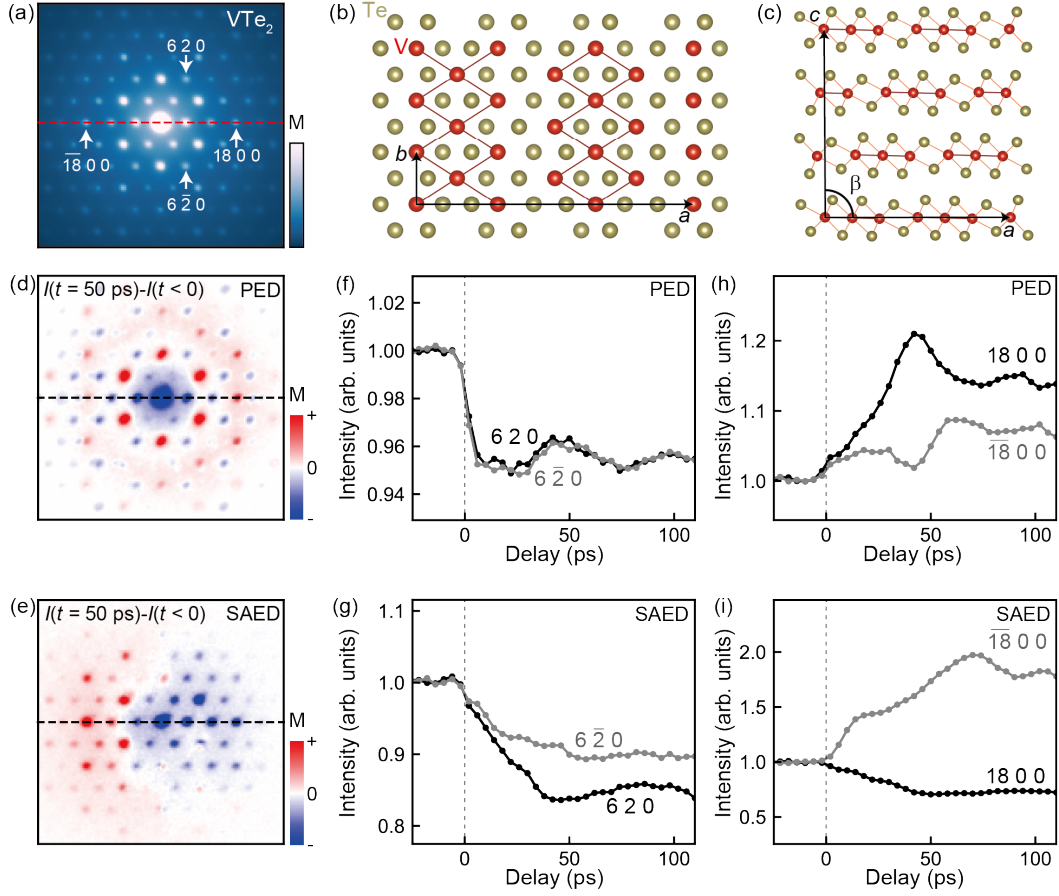

**Fig. 4. Ultrafast time-dependent PED intensity of VTe$_2$.** (a) PED pattern of VTe$_2$. (b,c) Crystal structure of VTe$_2$ at room temperature. (d,e) change in ultrafast PED and SAED intensity $\Delta I(t) = I(t = 50 \text{ ps}) - I(t < 0)$. (f-i) time-dependent diffraction intensity for 6 2 0, 6 $\bar{2}$ 0, 18 0 0, $\overline{18}$ 0 0 diffraction peaks obtained by ultrafast PED and SAED.

### 3.3. Crystal structure refinement

Based on the ultrafast PED intensity obtained by 4D-PED measurement, we further demonstrate the crystal structure refinement for the photoinduced state of VTe$_2$. The crystal structure refinement at $t < 0$ and $t = 6$ ps was performed by minimizing the so-called R-factor:

$$R = \frac{\sum_{hkl}\left|I_{\text{exp}}^{hkl} - I_{\text{calc}}^{hkl}\right|}{\sum_{hkl}\left|I_{\text{exp}}^{hkl}\right|}, \qquad (2)$$

where $hkl$ is a Miller index of diffraction peaks, $I_{\text{calc}}^{hkl}$ is the diffraction intensity obtained from the diffraction simulation. In Ref. [29], the R-factor is significantly decreased when the dynamical diffraction simulation is used as compared with that of kinematical diffraction. Therefore we employed the dynamical diffraction simulation to calculate $I_{\text{calc}}^{hkl}$. The diffraction peaks used to calculate the R factor in VTe$_2$ were limited to 72 points, that is, Bragg diffractions whose scattering vector was less than 8.0 Å$^{-1}$ because very weak diffraction intensities outside this region are difficult to precisely analyze. We fixed all lattice parameters that were reported in the equilibrium state because we did not observe significant changes in the position of Bragg peaks (i.e., length of *a* and *b* axes) nor the monoclinic angle $\beta$ as described in Sec. 3.4 within the first 6 ps. In addition, we fixed all atomic coordinates parallel to the optical axis because they do not affect the result of multislice simulation [42]. Under these assumptions, the parameters that can be determined for the crystal structure of VTe$_2$ under the space group of low-temperature C2/m are four atomic coordinates along *a* axis (*x*) as given in Table 1. All other fixed parameters are taken from Ref. [47]. For the *t* < 0 data, we optimized all these four parameters and the thickness simultaneously. The optimized thickness was 62 unitcells, i.e., 40.0 nm. For *t* = 6 ps data, we fixed the thickness and optimized the four parameters in Table 1. For the optimization, Bayesian optimization implemented in the Python scikit-optimize package was used. As a result of optimization, we achieved $R = 0.100$ and 0.095 for *t* < 0 and 6 ps data, which are sufficiently small values indicating the successful crystal structure refinement [30].

|  | $t < 0$ | $t = 6$ ps | LT (300 K) [47] | HT (500 K) [47] |
|---|---|---|---|---|
| $x_V$ | 0.160 | 0.168 | 0.147 | 0.167 |
| $x_{Te1}$ | 0.053 | 0.055 | 0.053 | 0.056 |
| $x_{Te2}$ | 0.223 | 0.221 | 0.224 | 0.222 |
| $x_{Te3}$ | 0.391 | 0.389 | 0.395 | 0.389 |

Table 1. The result of crystal structure refinement at t < 0 and 6 ps, as well as the temperature-dependent parameters reported in the literature [47].

Table 1 shows the result of our crystal structure refinement and at *t* < 0 and *t* = 6 ps, together with the equilibrium low-temperature (LT, 300 K) and high-temperature (HT, 500 K) values [47]. The crystal structure refinement at *t* < 0 and *t* = 6 ps indicates that the photoinduced atomic motion is primarily described by the vanadium atom, while the position of tellurium is not significantly changed. Similarly, the atomic coordinates of the vanadium atom mainly change between the LT and HT phases [47]. The refined crystal structures in Fig. 5 indeed show that the vanadium-vanadium chain structure is almost dissolved within the first 6 ps. These results are consistent with the previous ultrafast electron diffraction/microscopy measurements reporting the siginificant suppression of the superlattice peaks, and demonstrate the capability of 4D-PED for nonequilibrium crystal structure refinement.

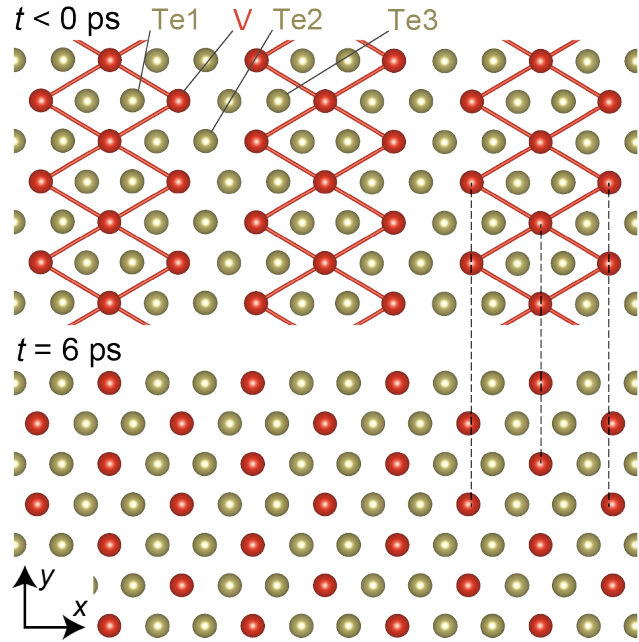

**Fig 5. Results of crystal refinement at $t < 0$ ps and $t = 6$ ps.** The black dashed line denotes position of vanadium atoms before photoexcitation.

### 3.4. Analysis of $\phi$ profile.

In addition to the diffraction intensity, the 4D-PED measurement further enables us to quantitatively evaluate the change in the reciprocal lattice vector parallel to the optical axis as described in Sec. 2.2. Fig. 6(a) shows time-dependent $\phi$ profile $I_\phi(\phi, t)$ of 6 0 0 diffraction intensity measured on $VTe_2$. $I_\phi$ has two peaks, at which 6 0 0 diffraction satisfies the Bragg condition. The distance between the two peaks decreases at $t > 50$ ps compared to $t = -22, 2,$ and $10$ ps. In the present case, such a time-dependent change in the peak positions should originate from a time dependence of $\beta$, which should affect $g_z^{600}$. Using Eq. 1, we calculated the time dependence of $\Delta\beta = \beta(t) - \beta(t < 0)$ under the assumption that the length of $a, b,$ and $c$ are not changed. Calculated $\Delta\beta$ in Fig. 6(b) monotonically changes from $t = 0$ to $t = 50$ ps by about $0.40°$. We note that this value is fairly close to the deviation of the original monoclinic angle (90.37°) of the room-temperature VTe$_2$ as compared to the high-temperature non-distorted phase (90°). Such a quantitative estimation of $\Delta\beta$ from electron diffraction has been difficult in previous studies [15,18].

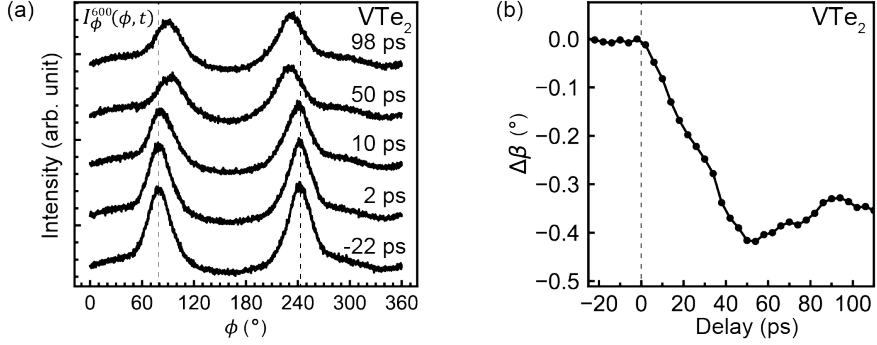

**Fig. 6 Time-dependent change in $\phi$ profile.** (a) Time dependence of $\phi$ profile. The dashed line denotes the peak positions at $t < 0$. (b) Time dependence of $\Delta\beta$.

## 4. Discussion

The application of 4D-PED is promising for future investigation of ultrafast phenomena. In the previous ultrafast electron diffraction and microscopy measurements, electron diffraction has been primarily used to estimate the change in lattice parameters and atomic coordinates. For example, changes in lattice parameters, e.g. acoustic phonons, have been investigated [14–20]. In addition, the changes in atomic coordinates have also been discussed in many photoinduced structural phase transition systems [19,21–28]. The application of the 4D-PED developed in this study will open a new pathway for the quantitative investigation of the crystal structure dynamics of these systems in the future. Further improvement of 4D-PED may be realized by utilizing advanced techniques such as aberration correction and energy-filtered PED [39], and automated electron diffraction tomography/3D electron diffraction [48–50]. Indeed, improvements in equilibrium structural analysis in various materials by these methods have been reported in the literature recently [51,52]. In addition, the integration of the PED technique with the recently developed 5D-STEM [12,13] will open a pathway to the quantitative spatiotemporal mapping of refined crystal structure in nm × ps scale.

5. conclusion

In conclusion, we developed a 4D-PED system and demonstrated its potential to detect quantitative changes in crystal structure. Combining an electron microscope, a pixelated camera, and a pump-probe method enables us to obtain time- and electron-incident-angle dependent four-dimensional electron diffraction patterns. We verified the 4D-PED system by measuring single crystalline silicon, where quantitative agreement between the experimental results and dynamical simulation was achieved. Further investigation of photoinduced crystal structure dynamics of $VTe_2$ reveals the capability of 4D-PED for the nonequilibrium crystal structure refinement. In addition, we showed and demonstrated that the change in reciprocal lattice vectors parallel to the optical axis can also be quantitatively evaluated from the time-dependent peak position in the $\phi$ profile. These results demonstrated the advantage of 4D-PED for the quantitative investigation of crystal structure dynamics, which will play an important role in the future investigation of ultrafast phenomena.

**Note**

The authors declare no competing interest.

**Acknowledgments**

We thank S. Akatsuka, N. Mitsuishi, H. Takahashi, and S. Ishiwata for the preparation of $VTe_2$ sample. This work was partially supported by Grant-in-Aid for Scientific Research (KAKENHI) (Grant No. 21K14488). J. Koga acknowledges the support from QSTEP at the University of Tokyo.

**CRediT authorship contribution statement**

**T. Shiratori:** Data curation, Formal analysis, Investigation, Resource, Software, Validation, Visualization, Writing-original draft. **J. Koga:** Data curation, Investigation, Resource, Funding acquisition, Software, Validation. **T. Shimojima:** Funding acquisition, Validation, Supervision. **K. Ishizaka:** Funding acquisition, Supervision, Validation, Writing-review & editing. **A. Nakamura:** Conceptualization, Formal analysis, Investigation, Project administration, Software, Supervision, Validation, Writing-review & editing.